\documentclass{aa}
\usepackage{txfonts}
\usepackage{graphicx}
\begin{document}

%

   \title{G0.087$-$0.087, a highly polarized flat spectrum filament \\
near the Galactic Centre
Arc\thanks{Based on observations with the Effelsberg 100-m telescope
operated by the Max-Planck-Institut f\"ur Radioastronomie (MPIfR), Bonn,
Germany}}

   \author{W. Reich}

   \offprints{W. Reich}

   \institute{
             Max-Planck-Institut f\"ur Radioastronomie, Auf dem H\"ugel 69,
             53121 Bonn, Germany\\
             email: wreich@mpifr-bonn.mpg.de }

   \date{Received 29 November 2002 / Accepted 31 January 2003}

\abstract{
32~GHz observations with the Effelsberg 100-m telescope revealed a
highly polarized Galactic Centre filament: G0.087$-$0.087, running
parallel to the well-known Arc structure. It has a similar flat
spectrum, but it is an order of magnitude weaker. G0.087$-$0.087 is
about $2\farcm 5$ or 6.2~pc long and located at the western boundary
of the expanding molecular cloud G0.11$-$0.11. This unusual cloud
is also visible in X-rays and known to interact on its eastern
periphery with the Arc. Acceleration of highly relativistic particles
at the surface of G0.11$-$0.11 seems necessary to explain the properties
of both structures. The new filament supports previous findings on the
existence of a strong poloidal magnetic field throughout the Galactic
Centre region.
\keywords {Galaxy: center -- radiation mechanisms:
non-thermal -- techniques: polarimetric} }

\titlerunning{G0.087$-$0.087, a flat spectrum filament near the Arc}
     \maketitle
%

\section{Introduction}

The Galactic Centre (GC) Arc is one of the most unusual radio
structures in the Galaxy. It is located at a projected distance of
about 30~pc from the Sgr~A complex and consists of long, thin, almost
straight filaments, running nearly perpendicular to the Galactic plane,
embedded in diffuse emission. The spectrum of the Arc is flat or
inverted up to 43~GHz (Reich et al.\ \cite{rei88}) and compatible with a
quasi-monoenergetic particle spectrum. This holds for its most intense
section of about 40~pc in length within the Galactic disk. Its spectrum
steepens drastically away from the plane, where the Arc can be followed
up to about 400~pc distance (Reich\ \cite{rei90}; Pohl et al.\
\cite{pohl92}). High-frequency observations beyond 10~GHz are required
to directly trace the magnetic field structure of the Arc, since large
rotation measures (RMs) exceeding several thousands $\rm rad\,m^{-2}$
are observed, and in addition strong depolarization is present.

Early Effelsberg 32~GHz measurements including linear polarization
(Reich\ \cite{rei90}; Lesch \& Reich\ \cite{lesch92}) revealed
percentage polarizations of about 50\% close to its intrinsic value of
about 60\% with the magnetic field running almost parallel to the Arc.
Its narrow filaments disappear at 43~GHz, while the flux density
dominating diffuse component is still visible (Sofue et al.\
\cite{sof92}). These observations agree with a magnetic field strength
in the milli--Gauss range for the filaments as proposed by Yusef-Zadeh
\& Morris (\cite{yus87}) to resist the pressure by interstellar clouds.
However, such strong magnetic fields limit the synchrotron lifetime to
a few hundred years, which is significantly below that of the diffuse
Arc component where the magnetic field strength is of the order of
$10^{-4}$ Gauss assuming equipartition (Reich\ \cite{rei90}).
Several particle acceleration scenarios have been proposed to explain
the unusual properties of the Arc. The detection of excessive 150~GHz
emission (Reich et al.\ \cite{rei99}) at the boundary of a massive
expanding molecular cloud (Tsuboi et al.\ \cite{tsu97}; Oka et al.\
\cite{Oka97}) has strengthened the idea of particle acceleration by
magnetic reconnection at the surface of that cloud.

The Arc is a unique structure, although deep low-frequency
observations revealed a number of similar although much fainter and
shorter nonthermal filaments (NTFs) in the GC area (LaRosa et al.\
\cite{la00}). The NTFs have in general steep spectra. Some of them show
a spectral steepening with frequency (Lang et al.\ \cite{la99}) or
a spectral index gradient along their length (LaRosa et al.\ \cite{la01}).
Like the Arc many NTFs are located close to molecular clouds. The NTFs
may represent fading relics of structures similar to the Arc, where
particle acceleration stopped or became inefficient. Strong magnetic
fields then quickly steepen the spectra.

In the course of a larger 32~GHz GC survey a weak filament,
G0.087$-$0.087, with properties like the Arc was identified, which is
described in this paper.

\section{Observations and data reduction}

The GC region was observed at 32~GHz with the Effelsberg 100-m telescope
in several clear nights between 2000 and 2002.  Details of the
performance of the telescope at that frequency are listed in
Table~\ref{tab:obs}. 3C286 served as the primary calibrator at 32~GHz
(Table~\ref{tab:obs}). A particular problem is the low elevation of the
GC region at Effelsberg, which limits observations within elevations
between $8\degr$ and about $11\degr$. This results in an observing time
of about three hours a day. To guarantee pointing and flux density
calibration at these low elevations all measurements were tied to the
peak flux density of Sgr~A (including Sgr~A* and Sgr~A West), which is
only partly resolved by the $26\arcsec$ beam. Its peak flux density for
the Effelsberg beam was measured as 9.7~Jy (Morsi \& Reich\
\cite{mor86}), which nicely fits to flux densities obtained from other
high-frequency observations (Tsuboi et al.\ \cite{tsu88}).

Polarization data were collected with a three-feed system (MOD~1)
installed in the secondary focus of the telescope. The two circular
components of each feed were connected to two cooled HEMT receivers
with 2~GHz bandwidth followed by an IF-polarimeter. The feeds were
aligned along azimuth--direction with spacings of $2\arcmin$, 4\farcm 3
and 6\farcm 3, which requires mapping along the azimuth--direction. The
complex extended emission in the GC region requires to make long scans of
$30\arcmin$ to $40\arcmin$ to restore all components properly.

MOD~1 also receives total intensities, but the sensitivity is low.
Since 1999 a second module (MOD~2) is available with a different
technical concept to measure total intensities only. The circularly
polarized components of different feeds were connected by waveguides
into a magic-T. The signal difference between two feeds was
obtained by a correlation of the magic-T outputs. The data were restored
into a single beam map using the Emerson et al. (\cite{emer79}) algorithm.
Adding all three spacings of MOD~2 a three times higher sensitivity for
total intensities was achieved than with MOD~1 (Table~\ref{tab:obs}).
However, at the low elevations of the GC atmospheric emission reduces
the nominal sensitivities significantly by extinction and increased
system noise. MOD~2 has the same feed separations as MOD~1 and was
installed with an offset of $153\arcsec$ in elevation parallel to MOD~1
with the same azimuth offsets. Pointing errors of a few arcsec and
small focal differences between individual coverages reduced the angular
resolution to about $27\arcsec$.

\begin{table}
  \caption{Basic observational parameters}
  \label{tab:obs}
  {\begin{tabular}{ll} \hline
  \noalign{\smallskip}
  HPBW & $26\arcsec$ \\
  Maximum aperture efficiency & 23\%  \\
  First sidelobe maxima/mean & $-$12~dB/$-$17~dB \\
  Instrumental polarization & $\le 1\% $ \\
  $\rm T_{B}/S~[K/Jy]$ & 1.8 \\
  \noalign{\smallskip}\hline
  \noalign{\smallskip}
  Nominal rms-noise/sec: &  \\
  MOD 1: I/PI & 12/2.4~mJy/beam\\
  MOD 2: I    & 4~mJy/beam \\
  \noalign{\smallskip}\hline
  \noalign{\smallskip}
  Main calibrator: 3C286: &\\
  S [Jy]  & 2.1 \\
  Percentage polarization [\%] & 13\\
  Polarization angle [$\degr$] & 33\\
  \noalign{\smallskip}\hline
  \end{tabular}}
\end{table}

\section{Results and discussion}

\begin{figure}[htb]
\centering
\includegraphics[bb = 66 138 378 437,width=8.8cm,clip=]{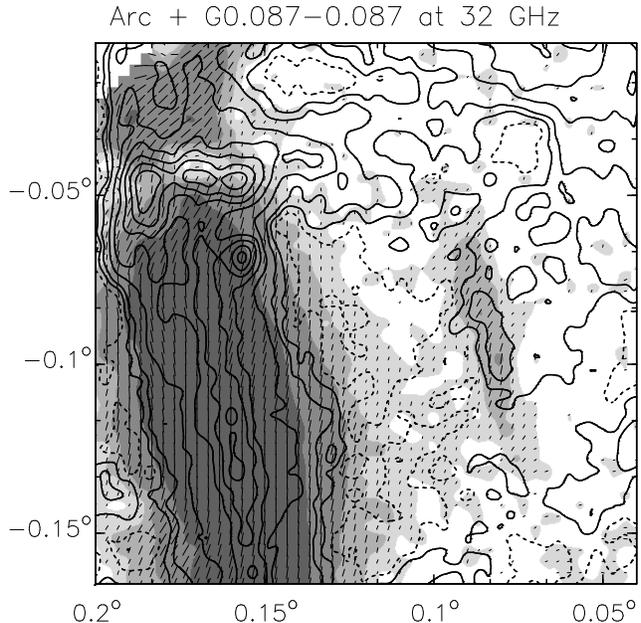}
\caption{G0.087$-$0.087 and a section of the Arc at 32~GHz at
$27\arcsec$ angular resolution. Total intensities are shown as contours
with a large-scale diffuse emission gradient removed. Contours run
from zero (dashed) up to 100~mJy/beam in steps of 50~mJy/beam and
beyond 200~mJy/beam in steps of 100~mJy/beam. Polarized emission is
overlaid in greyscale. Polarization bars are in B-field direction for
the case of negligible Faraday rotation. Their length is proportional
to polarized intensities up to 40~mJy/beam and constant above. Vectors
are shown for intensities exceeding 8~mJy/beam ($2\times$ rms-noise).
The polarized peak flux of G0.087$-$0.087 is about 40~mJy/beam and that
of the Arc is about 380~mJy/beam.}
   \label{rmlowres}
\end{figure}

\begin{figure}[htb]
\centering
\includegraphics[bb = 66 138 378 437,width=8.8cm,clip=]{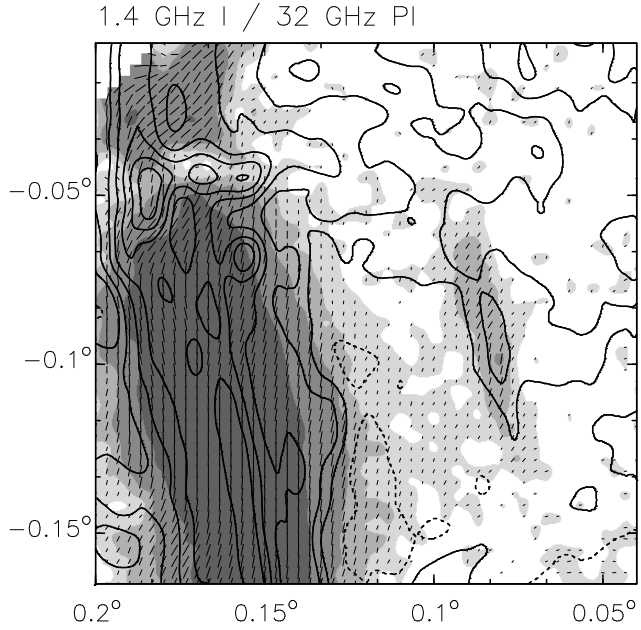}
\caption{G0.087$-$0.087 and a section of the Arc at 1.4~GHz (Yusef-Zadeh
\& Morris \cite{yus87}) shown as contours. The VLA 1.4~GHz total intensity
image is shown convolved
to $27\arcsec$ angular resolution. Effelsberg 32~GHz polarized intensities are
overlaid as in Fig.~\ref{rmlowres}. Total intensity contour levels are
the same as in Fig.~\ref{rmlowres}. }
   \label{lowres}
\end{figure}

In Fig.~\ref{rmlowres} the total intensity map of G0.087$-$0.087 at
32~GHz is shown including a section of the Arc. The rms--noise of the
total intensities was measured to be about 10~mJy/beam. Polarization
bars are superimposed. The rms--noise in polarized intensity is about
4~mJy/beam. G0.087$-$0.087 is outstanding in polarization, while its
total intensity is somewhat masked by large-scale emission. To enhance
the contrast large-scale total intensity structures were partly
removed by applying a background filtering technique (Sofue \& Reich\
\cite{sof79}). The length of the filamentary structure G0.087$-$0.087
is about $2\farcm 5$ corresponding to a projected length of about
6.2~pc for a GC distance of 8.5~kpc. The intensity along the filament's
ridge is about 70~mJy/beam with a percentage polarization of about 50\%.
G0.087$-$0.87 is much shorter and about eight times weaker than the
adjacent section of the Arc. Both structures run almost parallel
deviating by the same inclination angle of about $9\degr$ from
Galactic latitude direction. G0.087$-$0.087 indicates that the magnetic
field direction traced by the Arc extends for a larger area.

We have searched for counterparts of G0.087$-$0.087 in lower frequency
maps. Single-dish data, which all have lower angular resolution, show
no sign of G0.087$-$0.087 in total power or polarization (e.g.
Seiradakis et al.\ \cite{sei89}). This is not unexpected in view of the
dominating large-scale emission. Also polarized emission is not visible
due to strong depolarization in the GC direction, which even masks the
much stronger polarized emission from the Arc largely.

The polarization angle of G0.087$-$0.087 is in the range of $63\degr$ to
$80\degr$. In case of no Faraday rotation the magnetic field direction
would be inclined by about $-19\degr$ or $-36\degr$ in respect to the
filament's axis. However, in the most likely case that the magnetic
field is parallel to the filament the observed angle deviations imply
RMs of the order of $\rm -3800~rad\,m^{-2}$ to $\rm -7200~rad\,m^{-2}$.
With the same argument the RMs along the Arc are between $\rm
-2400~rad\,m^{-2}$ and $\rm -3400~rad\,m^{-2}$. Such high RMs are
unusual for Galactic emission structures, but rather similar to those
reported previously for the Arc structure (Tsuboi et al.\ \cite{tsu87};
Sofue et al.\ \cite{sof87}; Yusef-Zadeh \& Morris\ \cite{yus87b}).

\begin{figure}[htb]
\includegraphics[bb = 18 14 720 371,width=8.8cm,clip=]{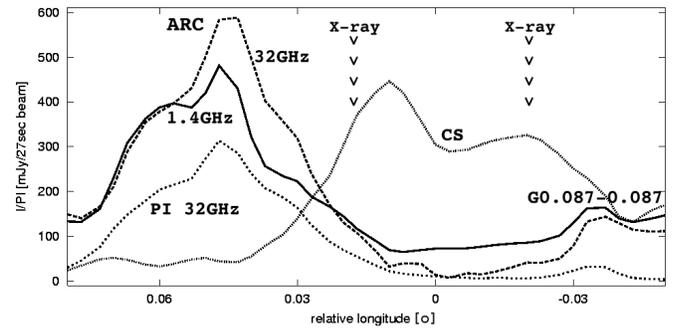}
\caption{Cuts across the Arc and G0.087$-$0.087 taken from the maps as
shown in Fig.~\ref{rmlowres}, Fig.~\ref{lowres} and
Fig.~\ref{csemission}, but rotated by $9\degr$ inclination of the Arc
and G0.087$-$0.087 against Galactic latitude. The integrated CS emission
from G0.11$-$0.11 is shown on a relative intensity scale. The position
of the X-ray filaments running parallel to the Arc and G0.087$-$0.087
as observed by Yusef-Zadeh et al. (\cite{yus02}) is
indicated.}
   \label{cuts}
\end{figure}

Weak filamentary total intensity emission of G0.087$-$0.087 is
identified in a 1.4~GHz VLA map published by Yusef-Zadeh \& Morris
(\cite{yus87}), where numerous weak filamentary structures are
visible. This interferometer map does not include large-scale emission.
Many filaments of similar strength are noted on high-resolution VLA
maps at low frequencies (e.g. LaRosa et al.\ \cite{la00}), but none of
them has a counterpart at 32~GHz like G0.087$-$0.087 as expected for
steep-spectrum emission. In Fig.~\ref{lowres} we show 1.4~GHz total
intensities from the VLA (Yusef-Zadeh \& Morris\ \cite{yus87}) with
32~GHz polarized intensities superimposed for the same area as in
Fig.~\ref{rmlowres} and at the same angular resolution of $27\arcsec$.
The wide frequency interval and the rather similar total intensity
levels in Fig.~\ref{rmlowres} and Fig.~\ref{lowres} indicate rather
flat or slightly inverted spectra. From orthogonal cuts across the
filaments (shown in Fig.~\ref{cuts}) a spectral index of $\rm \alpha
\sim 0.15~(S \sim \nu^{\alpha})$ is calculated for G0.087$-$0.087. Cuts
across the Arc give a spectral index of $\rm \alpha \sim 0.10$ for its
maximum emission. These spectral indices are rather similar to those
reported by Reich et al. (\cite{rei88}) from a multifrequency study
of the Arc between 863~MHz and 43~GHz at $1\farcm 2$ angular
resolution. The missing large-scale structures in the VLA map seem to have
little influence on the spectral indices derived for distinct features like
the Arc and G0.087$-$0.087. A rather conservative estimate for possible
scaling errors of up to 10\% for each map result in a spectral index
change of about $\rm \delta \alpha \le 0.06$. The similarity of
spectral indices for the Arc and G0.087$-$0.87 suggest that they were
created in a similar process. Both structures are different from the
majority of nonthermal filamentary structures in the GC region with
steep spectra, which are too faint to show up in the present 32~GHz map.

\begin{figure}[htb]
\centering
\includegraphics[bb = 66 138 368 437,width=8.8cm,clip=]{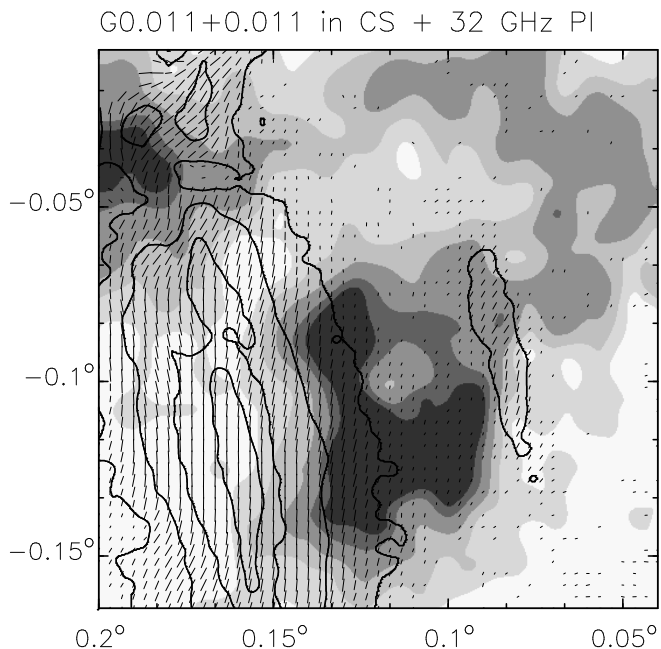}
\caption{Integrated CS (J=1--0) emission (15~km/s~$\le v \le 45$~km/s)
of the molecular cloud G0.11$-$0.11 (gray scale) with superimposed
contours of polarized emission at 32~GHz. Contours are shown at 20,
100, 200 and 300~mJy/beam.
Polarization bars are shown as in Fig.~\ref{rmlowres} and
Fig.~\ref{lowres}. }
  \label{csemission}
\end{figure}

The apparent interaction of the massive molecular cloud G0.11$-$0.11
with the Arc at its eastern periphery was discussed by Tsuboi et al.
(\cite{tsu97}). The cloud has a size of about 7.5~pc and a mass of $\rm
(2.0-3.6) \times 10^{5} M_{o}$. It expands and rotates with 20~km/s
and 4~km/s, respectively. Reich et al. (\cite{rei99}) rediscussed this
interaction in view of new 150~GHz observations with the Nobeyama 45-m
telescope, which show enhanced emission across the interacting areas.
This excessive high-frequency emission traces the region of particle
acceleration at the surface of G0.11$-$0.11. Magnetic reconnection is
the likely acceleration process. G0.087$-$0.087 is -- like the Arc --
seen towards the periphery of the molecular cloud, but on its western
side. It is rather suggestive to assume that the same interaction
process as for the Arc takes place. However, the much lower intensity
of the G0.087$-$0.087 filament suggests that the acceleration process is
less efficient compared to the Arc. Figure~\ref{csemission} displays
G0.087$-$0.087 and the Arc at the western and eastern periphery of the
molecular cloud G0.11$-$0.11.

G0.087$-$0.087 is barely resolved at 32~GHz. The unconvolved 1.4~GHz
VLA total intensity data with an angular resolution of $17\farcs
1\times 16\farcs 4$ limit the intrinsic width of G0.087$-$0.087 to about
$20\arcsec$. The polarized intensity distribution (Fig.~\ref{rmlowres}
and Fig.~\ref{lowres}) indicates weak diffuse emission slightly above
the noise level extending across G0.11$-$0.11. Thus it may also be
possible that the molecular cloud is surrounded by synchrotron emission
forming a thin torus with a thickness of about 1~pc. In such a case
G0.087$-$0.087 results from an edge-on view of the torus. Particle
acceleration takes place at a larger area on the surface of the
molecular cloud, where it is expanding in orthogonal direction to the
ambient magnetic field direction. In this scenario the Arc is the
exceptional structure, where additional conditions like magnetic field
compression are needed to make the particle acceleration process more
efficient than elsewhere on the surface of G0.11$-$0.11. The excessive
brightness of the Arc compared to G0.087$-$0.087 also indicates a larger
magnetic field strength.

It has been shown observationally (Sofue et al.\ \cite{sof92}, Reich et
al.\ \cite{rei99}) that the Arc is a rather short-living structure not
exceeding a lifetime of a few thousand years. Reich et al.
(\cite{rei99}) argue for a non-pressure equilibrium between the Arc and
G0.11$-$0.11 to avoid magnetic field strenghts exceeding 1~mGauss, which
seem not to be compatible with the minimum lifetime of the Arc given by
its physical length. A weaker magnetic field for G0.087$-$0.087 implies
a longer lifetime and a spectral turn-over will be shifted towards
higher frequencies when compared to the Arc. However, its intrinsic
faintness makes G0.087$-$0.087 a difficult object to observe and its
physical parameters are therefore not easy to constrain.

G0.11$-$0.11 is a very unusual molecular cloud in the GC region with a
high kinetic temperature (Oka et al.\ \cite{Oka97}). CHANDRA
observations of the G0.11$-$0.11 area revealed diffuse X-ray emission
distributed across the cloud with enhanced filamentary emission along
the edges of the molecular cloud in the direction parallel to the Arc
and G0.087$-$0.087 (Yusef-Zadeh et al.\ \cite{yus02}, their Figs.~1 and 2).
The X-ray filament near the Arc is rather strong, a weaker one is clearly
visible towards G0.087$-$0.087, as indicated in Fig.~\ref{cuts}. These
observations further support the interaction scenario between the Arc,
G0.087$-$0.087 and G0.11$-$0.11. Yusef-Zadeh et al. (\cite{yus02})
proposed that relativistic electrons, which likely are from the
acceleration process, provide the heating of the molecular cloud. The
unidentified EGRET source 3J1746$-$2851 (Hartman et al.\ \cite{hart99})
might also result from that interaction. This supports the previous
suggestion of Pohl (\cite{pohl97}) that the GC $\gamma$-ray source
coincides with the Arc rather than with the Sgr~A complex.

\section{Conclusions}

A new filament G0.087$-$0.087 with rather similar spectral and
polarization characteristics compared to the GC Arc has been
identified. It runs parallel to the Arc and is located at the western
periphery of the same molecular cloud (G0.11$-$0.11) believed to be in
interaction with the Arc. The same particle acceleration process,
although of lower efficiency, seems to take place at G0.087$-$0.087. It
may be a single filament or results from a faint emission torus
surrounding G0.11$-$0.11 seen edge-on. The orientation of the Arc and
of G0.087$-$0.087 with the same inclination of about $9\degr$
supports the idea of the existence of a large scale poloidal field
throughout the GC.

\begin{acknowledgements}
I like to thank Ute Teuber for her long lasting successful work on the
32~GHz Effelsberg receivers.
\end{acknowledgements}

\end{document}